# BIG DATA AND LEARNING ANALYTICS IN HIGHER EDUCATION
*Demystifying Variety, Acquisition, Storage, NLP and Analytics*

Amal S. Alblawi
Information Systems Department, Shaqra University,
Riyadh, Saudi Arabia

Ahmad A. Alhamed
Information Systems Department, King Saud University,
Riyadh, Saudi Arabia

*Abstract*—Different sectors have sought to take advantage of opportunities to invest in big data analytics and Natural language processing, in order to improve their productivity and competitiveness. Current challenges facing the higher education sector include a rapidly changing and evolving environment, which necessitates the development of new ways of thinking. Interest has therefore increased in analytics as part of the solution to many issues in higher education, including rate of student attrition and learner support. This study provides a comprehensive discussion on big data, learning analytics and use of NLP in higher education. In addition, it introduces an integrated learning analytics solution leveraging a distributed technology system capable of supporting academic authorities and advisors at educational institutions in making decisions concerning individual students.

## I. INTRODUCTION

There has been increasing interest in analytics as part of the solution to many of the issues experienced by the higher education sector, e.g. student attrition and learner support[1]. The major objective of learning analytics is concerned with creation of beneficial information for data-driven decision-making. This can be achieved by the analysis of Big Data collected from various learning sources[2]. However, the greatest challenge in higher education is to determine how data is captured, processed, stored, presented, and used for the benefit of tomorrow's outcomes [3]. One of the crucial issues faced by higher education institutions consists of constant (or increasing) rate of student dropouts or transfers[4, 5]. Till date, analytics in the high education sector have been limited to traditional Business Intelligence (BI), using traditional students' log data to make decisions and drive action. This has proven to be largely ineffective in improving student's retention and graduation, as traditional BI tools being incapable of handling additional factors (i.e. unstructured data), leading to a potential influence on student performance. Universities have recently begun to use Big Data analytics to address the urgent issue of student retention. The most common application used by universities consists of early warning systems, which rely on student performance data to predict and detect students at risk of failure. However, these pilot implementations of early warning systems adopt a limited number of analytic techniques and data features. The majority of current systems focus on a single outcome, with many relying on online educational data. This data may provide only a proportion of the potential insight into a student's engagement with a university. The education sector benefits from valuable research outcomes from previous studies on the uses of Big Data. However, such existing work lacks in depth insight into Big Data system architectures in the education sector for applied multi-analysis, or data collection and additionally, we are using NLP solutions. Thus, the published work focusing on the factors influencing student success, failure, or dropout rates is scarce. At the same time, to the best of our knowledge there is no technical design adopted to incorporate such factors, or consider the key types of learning analytic techniques and how they apply to a combination to determining student performance or dropout rates. Therefore, addressing such issues can provide an extension to, and an improvement of, the decision support systems currently employed by education management, thus improving its usefulness and practical applicability. In this model, NLP is merged with data analysis. The benefit of merging NLP solutions with other forms of data analysis is the ability to accurately analyse increasing amounts of unstructured data, such as student feedback. This should provide more accurate insights into human behaviour, especially when combined with other structured data. Initially, it is necessary to check the significance of both the datasets. If the target variable correlates with the text data, the result of the text data can be combined with the main data to improve the accuracy of the results. To summarize, these methods could be used in future models to improve the overall success rate. It is important that big data can be handled on different platforms. In general, Apache Spark is used to manage massive amounts of data and to provide real-time analytic power. In this model, we used Apache Spark for fast computations and for managing and processing big data. The second section of this paper provides a literature review. The third section presents the proposed system, including data collection, factors affecting student performance and dropout, NLP solutions and proposed Data Points for Decision Support Systems. The fourth section presents the experiment and the fifth section is the conclusion.

## II. REVIEW OF LITERATURE

In this section studies related to Big Data architecture, the application of Big Data Analytics related to learning, and Natural language processing are reviewed.

### A. Big Data Architecture

The opportunity of taking advantage of emergent Big Data and NLP solutions is desirable across various domains. This is studied by many researcher's scholars and it seeks to provide a general definition and framework for Big Data ecosystem. For instance, in[6]the authors provide an overview of the main characteristics of Big Data computing systems discuss the main layers in Big Data architecture relative to the traditional database model. In addition, the authors identify and classify gaps in the area of Big Data research, stating that there is need for research in several areas related to Big Data computing, such as data organization, decision-making, domain specific tools and platform tools. This can create a next-generation Big Data infrastructure that enables users to extract maximum benefit from the large amounts of data available. In the context of learning, this research aims to utilize multiple sources of student data in the higher education setting, describing suitable methods for its data acquisition and integration into a next-generation Big Data infrastructure.

### B. Application of Big Data Analytics in Learning

In the aim of analyzing massive amounts of data in an effective manner, Phoenix University used Hadoop solutions to deal with the issue of massive volume of data, such as the data generated by discussion forum databases. The main two issues that Hadoop solved at Phoenix University were as follows:

1- The digestion of large datasets (such as web usage logs, discussion form data)

2- Data analysis for unstructured data by running cycles of queries.

After the Hadoop processes from where the data and required information were derived, the summarized information was sent back to IDR for further analytics. Multiple analytics projects can be undertaken using the Phoenix University IDR. One such project predicts the retention rate of students by using available data such as schedules, grades, content usage, and students' demographic data[7].

### C. Natural language processing

Given the importance of obtaining real-time feedback from students to improve educational outcomes, we found that many previous studies have tried to analyse student feedback and extract student sentiment using different techniques and tools. In[8], machine learning techniques, preprocessing levels and the use of neutral class were used to analyse real-time student feedback. It was found that preprocessing the data improved accuracy by 20%. Another approach was used in [9] to analyse students' feedback using a lexicon-based approach. In fact, using a lexicon-based approach is not the best method of sentiment analysis, since some important information may be missed. To address this problem, Stanford CoreNLP, which was used in this study, utilizes a deep learning model to compute sentiment by taking into account how some single words can change the meaning of phrases[10]. Although the Stanford CoreNLP tool provides highly accurate results, it suffers from a slow processing time; adding additional annotators also makes it considerably slower. The Stanford CoreNLP tool was therefore run on a distributed framework, such as Spark, to improve the processing speed to bring it closer to real time. In[11], the study results showed that using Stanford CoreNLP on the Spark framework could increase processing speeds by up to 98%.

## III. PROPOSED SYSTEM

This section outlines the proposed solution by exploring non-traditional data points concerning student's data that, when analyzed, could lead to more accurate decision-making insights for higher academic authorities. At the same time, it shows the approach employed to collect the data from multiple resources for analysis.

### A. Factors Affecting Student Performance and Dropout

Academic performance engagement metrics are considered as an effective means of predicting students' success. However, there are a number of further important factors which can influence student's success, e.g. social integration. A number of studies have confirmed that social adjustment plays a critical role in student motivation, revealing that students with broader, well-connected networks were more likely to persist[5, 12]. Also, attitudes of students can have a positive impact on behavior such as intrinsic motivation and curiosity which trigger the learning actions. In contrary negative attitude of the student towards any learning environment component, may lead to boredom, anxiety, or stress, which may result in reducing the student's learning ability and eventually to student withdrawal. The review of literature enables factors influencing student retention to be categorized as follows: 1) Academic Integration (i.e. student GPA; marks; satisfaction with academic experience; and interest in classes and programs). 2) Social Integration (i.e. relationship with other students; impact of peer group; social/peer support; and extracurricular activities). 3) Institutional Commitment (i.e. funding; physical facilities; academic support; technological support; hands-on learning experiences; and academic advice). 4) Out-of-institution factors (i.e. finance; health; external social circles; and lifestyle)

### B. Proposed Data Points for Proposed System

The data types found in Table 1 facilitate the identification of all possible factors impacting student dropout and retention rates. In many cases, such data includes semi-structured and unstructured data, and requires a non-traditional data management system. Furthermore, the current ability to collect this form of data implies that it needs to be considered as an important component in the analysis of student performance to predict the rate of retention. The volume of data remains expressed in levels of gigabytes, however, it reveals a considerable variety. This indicates the potential of Big Data, and the need to tailor the solution to suit Big Data analytics.

TABLE 1. PROPOSED DATA POINTS

| Data Type | Examples of Data Variables |
|---|---|
| A) Student Logs | Name; Age; Gender; Location; Previous School; School graduation marks; |
| B) Student's Performance Statistics | Subject-wise internal assessment marks; Mid-term grades; Annual Examination grades; Laboratory marks; Project marks. |
| C) Student Engagement Metrics | Daily attendance; Seminar participation; Group study participation; Workshop attendance; Feedbacks/Reviews. |
| D) Student's Online Learning Engagement | LMS course list; LMS Login/Logout timestamp; LMS duration/day; LMS examination marks; LMS Modules completed; |
| E) Past student Achievement | Student winners; students' marks; student extra-curricular awards; student dropout rate; |
| F) Student's Social Network | Student's study group; Student's circle of friends. |
| G) Student's Extra Curricular Activities | Student's membership of clubs; Student's attendance of clubs; Student's participation in competitions. |
| H) Student's Health Background | Is he/she disabled? Does he/she have any chronic diseases? |
| I) Student's Financial Background | Annual household income; Did the student have a loan? Fee payment delay record; Does the student have a scholarship? |

*C. Proposed Integrated Analytics Model*

This framework seeks to support students' progress and graduation, this part of the research is aimed to utilize different types of data, by passing them through different analysis and then feeding the results into a master analytic model. By this multiple attributes and features can be involved in order to predict student performance and discover the factor that impacts student performance. An integrated analytics model is proposed (as opposed to a single predictive model) because the data is quite varied and possess its own unique characteristics. Existing university infrastructure can assist in the collection of such data in a structured format. However, such systems lack the capacity to run complex analytics on such varied data. The variety of the data investigated (i.e. free-form text, social graphs, and normal operational data), along with complex analytics, requires the use of a non-traditional data management system. The distributed platforms have arisen specifically in response to the failure of a traditional data warehouse to handle (a relatively low cost) large, complex data sets and deliver the output/response in a few seconds. Therefore, the current solution is to employ a distributed system, is robust and capable of handling a variety of data and scales in a rapid manner. The following section outlines the proposed Big Data architecture.

*D. Distributed Architecture Components of the Data Pipeline*

The components of the proposed solution architecture have been selected based on the previous study to address: (1) the nature of the data; (2) tools employed in data collection; and (3) the performed analytics. "Fig. 1" demonstrates the proposed architecture.

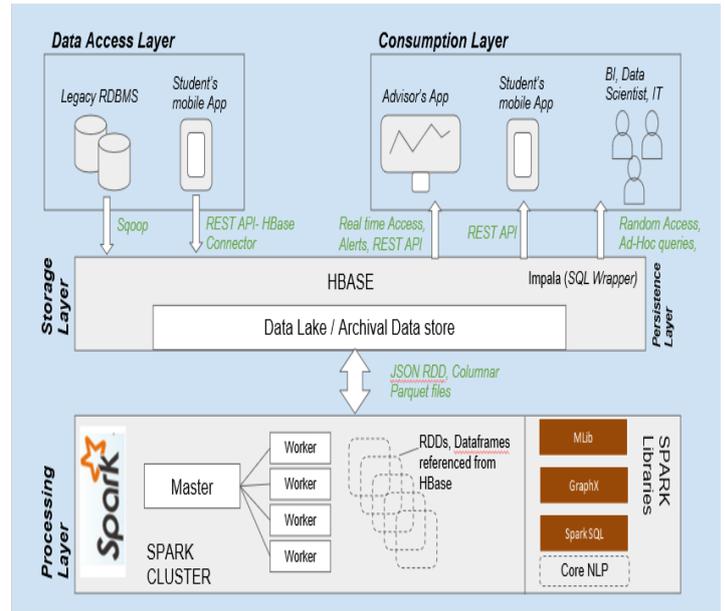

Fig. 1. Proposed Architecture.

*1) Data Access Layer*

The Data Access Layer comprises all the data sources the processing engine required to access. The available data sources are as follows:

- The university legacy database systems, e.g. student logs; student records; and historical data.

- The student mobile app, i.e. the app capable of generating data based on a student's activity.

Firstly, a data dump is required prior to pushing it into the processing engine (Spark). H-Base is considered appropriate for this action, as it: (1) stores data in its raw format; (2) provides real-time access; and (3) is convenient for batch processing. The following tools are proposed for data ingestion into H-Base:

- Sqoop: appropriate connectors are essential to ingest data into Spark from structured databases, H-Base. Sqoop can be used to convert stored structured data in SQL databases into a distributed file format, capable of being ingested by HBase[13]. Thus, all the university's databases can be connected to Sqoop, which can also be scheduled to ingest data at regular intervals, in order to capture any data updates.

- REST API Connector for H-Base: this enables the direct storing of data from the app in both a structured and unstructured format, as it arrives. The API can be programmed to fetch data as it changes in the app.

*2) Storage Layer*

The storage layer comprises HBase and its file system, HDFS. An Impala wrapper on HBase allows BI Analysts to run SQL queries on the data stored in HBase. A total of approximately 120 GB of data can be anticipated, i.e. based on an assumption of approximately 2MB data per student in the university (records, LMS data, app data, etc.). In the case of approximately 60,000 students in a university, this results in approximately 120,000MB (~120GB). This HBase therefore requires a minimum storage capacity of 120GB. If fault-tolerance is required, there will be a need to replicate the data, which will increase the space required. This will, in turn, depend on the number of times the data needs to be replicated by configure Replication parameter.

The collected data is now stored in a single location as Parquet files (columnar), in order to save space and facilitate distributed/random access. This distributed storage enables access to any data variable, as it comes from one large table, and further enables the processing engine to run iterative machine learning queries.

*3) Processing Layer*

Spark forms the processing layer in which all analysis will take place. All the data is in the form of Resilient Distributed Data (RDD) when Spark is used. The RDDs were created by referencing data stored in HBase which is used as external storage.

*a) Sentiment Analysis on Spark*

Thanks to programming language APIs provided by Spark, in addition to advance open source NLP libraries (such as Core NLP), many of the NLP functionalities can be imported to our system. Although the Stanford CoreNLP tool provides highly accurate results, it suffers from a slow processing time; adding additional annotators also makes it considerably slower. The Stanford CoreNLP tool was therefore run on a distributed framework, such as Spark, to improve the processing speed to and to bring it closer to real time.

*b) Predictive Analytics on Spark*

In this stage of the integrated analytics model, all the features (consisting of both computed features from the unstructured data analytics and original features) were brought to a final RDD. Generally, all predictive modelling procedures are done via a process of ensemble modelling; i.e. the use of different algorithms to test more accurate results. The Random Forest process can solve both types of prediction problems (Classification and Regression) via a process of numeric prediction. Accordingly, it was decided to build both predictive models, namely: Student Performance (regression) and Student Dropouts (binary classification). The (MLib), which is a spark's machine learning library, supports the use of Random Forest process for binary and multiclass classification and for regression. It uses both continuous and categorical features.

## IV. EXPERIMENT

The purpose of this experiment (pilot implementation) is to show that using the proposed integrated learning analytics technique is more efficient than a traditional single predictive model, especially if the data is fairly varied and has unique characteristics. The ability to handle and analyse various types of data (structured, semi-structured or unstructured) is one of the most important characteristics of Big Data analytic techniques. Thus, in this experiment we used multiple data types to examine students' performance. Using the proposed system, we built two predictive models. The first model was built using the traditional structured data to predict the students' final performance. The second model was built using the same data in addition to the students' sentiment scores. The students' sentiment scores were calculated by analysing students' textual feedback using the Stanford CoreNLP Natural Language Processing Toolkit. Finally, we drew our conclusions by comparing the accuracy of each predictive model. This section describes the steps and the results of this experiment.

*A. Data*

The programming course is one of the most challenging courses for first-year students. We thus chose data from the first programming course at the computer science college for this experiment. The data set consisted of:

*1) Structured data*

Students' detailed semester marks in the 111 CS course were collected over one year. The subjects consisted of 500 students enrolled at the university. The following information was collected for each participant: (Student ID, GPA, Major, Passed Hours, absence rate, Quiz (5), MID1 (15)MID2 (20),Tutorial (2), Homework (3), Lab Total (10), Final Lab (5)Final exam, Total: 100%, Grade, Status).

*2) Unstructured data (text data)*

The students were asked to write about five sentences stating their opinion of the programming course (111CS). The students' opinions were collected using a Google form. Out of the 500 participants, only 126 students completed the questionnaire. The questionnaire contained one field, the student ID, and three open questions to motivate students to write more sentences.

*B. Experiment Steps*

*1) Model 1: Predict student performance without sentiment score feature.*

Firstly, we built a random forest model to predict the student's final performance. At this stage, we used only the structural features, without including the sentiment score. This was done by following the steps below:

Data were transferred to HBase in preparation for processing by the spark engine. The students' marks stored in the MySQL database were imported to Hbase using sqoop.

As the project specified many libraries, the SBT build tool was used to manage and compile all the source files. After adding the required libraries into the sbtfile (StanfordNLP, Hbase, scala, spark core, Spark-Hbase connector, spark machine learning libraries), spark was configured using Scala. The feature values and target values in the function PassengerDataToLP() were then specified.

Target Variable: *Total_100 (here, only students' final performance is taken based on user requirements; the target variable and algorithm can also be changed).*

Input variables: *absence_rate, Quiz_5, MID1_15, MID2_20, Tutorial_2, Homework_3, Lecture_Total_45, Lab_Total_10, Final_Lab_5, semester.*

The data were split into two parts, one for training and another for testing. The random forest model was trained by specifying the parameters, such as the following categorical features: info, num Trees, feature Subset Strategy, gini, max depth and max bins. After training the model, the test data were predicted, and errors are found. The accuracy of the model with the regression evaluation package is maintained because the random forest regression model was used. The regression evolution contains only five evolution metrics: MSE, RMSE, MAE, R-Squared and Explained Variance. R-squared was chosen for the evaluation of this model.

*2) Model 2: Calculate the students' sentiment score, through their text feedback about the 101CS course using the Stanford Core NLP Natural Language Processing Toolkit.*

We read the text data from HBase and stored it in RDD. The mapping was then applied to RDD to process the text data. For processing the text, Scala API was used for the Stanford Core NLP toolkit. The Scala code creates an instance of StanfordCoreNLP, which internally constructs a pipeline that turns a text into an annotation object, and returns various analysed linguistic forms. Using the properties object, we were able to specify what annotators to run. By running the pipeline on the input text, we were able to obtain all the sentence annotation results from each annotation object. For each sentence annotation in the sentence annotation, we created a tuple of sentence text and sentiment value. Finally, each word score was calculated at every leaf of the tree and a tuple of the sentence text and sentiment score was returned. The Stanford-core NLP sentiment models gave sentiment values on a 5-point scale (0-4) of 1 (0.1 to 1.99) = negative, 2 (2 to 2.99) = neutral, (3 to 4) = positive. The final result of the sentiment was generated in the list format from all the sentences in each student's feedback. The main method gave the average sentiment score of all the sentences in a row because more than one sentence in a row or key was obtained. "Fig. 2" shows the sample output of the students' sentiment score model.

| Student_ID | Sentiment_Score | Sentiment |
|---|---|---|
| 2167283345 | 3 | Positive |
| 2171003565 | 3 | Positive |
| 2171012645 | 1 | Negative |
| 2176000465 | 3 | Positive |
| 2176000935 | 3 | Positive |
| 2176001455 | 3 | Positive |
| 2176002440 | 2 | Neutral |
| 2176003825 | 3 | Positive |
| 2176007410 | 2 | Neutral |
| 2176009150 | 3 | Positive |
| 2176012755 | 2 | Neutral |
| 2176014325 | 1 | Negative |
| 2176014700 | 1 | Negative |
| 2180000015 | 1 | Negative |
| 2180617295 | 2.5 | Neutral |
| 2181000145 | 2 | Neutral |
| 2181000525 | 4 | Positive |

Fig. 2. SENTIMENT SCORE OF EACH STUDENT ID

*3) Model 3: Prediction of Student Performance after adding the Sentiment Score as an additional feature to those used in Model 1.*

The previous model created a new feature for students' data, representing their sentiment by transferring their feedback in natural language to a sentiment score. This enabled the derived sentiment score to be added to other student features. We saved the results of the previous model (the ID and sentiment score for each student) in the student database before adding it to other student metrics. To test the hypothesis that adding new features derived from unstructured data can improve the results of the predictive model, a random forest model was built to predict students' final performance using the same features as in Model 1 in addition to the sentiment score. We then Compare the evaluation metrics for Model 1 with the evaluation metrics for Model 2.

*C. Experiment Results and Discussion :*

The final accuracy of the algorithm was calculated using a separate dataset that was not used in the cross-validation (usually called a test set). The following table shows how the predictive model was improved by adding the sentiment score as a feature in Model 3.

TABLE 2: THE ACCURACY EVALUATION FOR MODEL1 AND MODEL3

| Evaluation Metrics | Model 1 | Model 3 | Improvement |
|---|---|---|---|
| R-squared | 0.79 | 0.89 | 0.10 (10%) |

As shown in Table 2, the accuracy of Model 3 is integrated predictive model and includes the features derived from unstructured data, improved by 10% of accuracy compared to Model 1, traditional single predictive model.

The results of this experiment provide evidence of the importance of unstructured data. Specifically, that structured data extracted from unstructured data has been proven to have a good effect on the predictive model's accuracy. This indicates that there are genuine indicators within textual student feedback that can be extracted and used as important predictors of student performance. This finding is consistent with a number of recent studies discussing students' Sentiment analysis (e.g. [9], [8] and [14]).

As in this experiment, a number of other studies utilized Apache Spark as a platform for processing unstructured data. All these studies [11, 15] acknowledge the high and rapid performance of Spark, which can be used to manage massive amounts of data and provide real-time analytical power.

V. CONCLUSIONS

Building up a holistic picture of student progress and taking sentiment into account alongside other personality factors produces more accurate results regarding the prediction of student performance. This paper focused on the diverse sources of data that are available for students in higher education. It further discussed the employment of both traditional and non-traditional data points concerning students in order to establish insights into the key issues influencing decision-making within the domain of higher education, for example student retention. In addition, this paper has outlined the techniques employed to collect data from multiple sources. This should produce an integrated learning analytics solution leveraging a distributed technology system capable of supporting academic authorities and advisors at educational institutions in making decisions concerning student retention rates and methods to improve their performance. Moreover, in the last section of this paper, the experiment results showed that our hypothesis was correct, since the integrated predictive model including features derived from unstructured data showed a 10% improvement in the accuracy of results compare with the traditional single predictive model.

This proposed solution for data collection, storage and analysis offers potential benefits for future development activities in the learning analytics field.